\documentclass[twocolumn]{aastex631}
\usepackage{amsmath}
\usepackage{mathtools}
\usepackage{color}
\usepackage{hyperref}

\newcommand{\specnotation}[2]{\ensuremath{\rm #1 \, {\scriptstyle #2}}}
\newcommand{\molH}{\ensuremath{\rm H_2}}
\newcommand{\atomH}{\specnotation{H}{I}}
\newcommand{\degree}{\ensuremath{^\circ}}
\newcommand{\sknew}{Sk-70$\degree$32}
\newcommand{\skold}{Sk-68$\degree$82}
\newcommand{\kms}{km s$^{-1}$}
\newcommand{\cmmt}{cm$^{-2}$}
\newcommand{\mcmmt}{\text{cm}^{-2}}
\newcommand{\dens}{cm$^{-2}$}

\newcommand{\rstwv}{\lambda_r}
\newcommand{\NJ}[1]{N_{#1}}
\newcommand{\ltNJ}[1]{\lten N_{#1}}
\newcommand{\lten}{{\rm log}_{10}}
\newcommand{\fmolH}{\ensuremath{f({\rm H_2}})}
\newcommand{\NmolH}{\ensuremath{N_{\rm H_2}}}
\newcommand{\NatomH}{\ensuremath{N_{\rm H \, I}}}
\newcommand{\nmolH}{\ensuremath{n_{\rm H_2}}}
\newcommand{\natomH}{\ensuremath{n_{\rm H \, I}}}
\newcommand{\nH}{\ensuremath{n_{\rm H}}}
\newcommand{\NH}{\ensuremath{N_{\rm H}}}
\newcommand{\rng}{\text{--}}
\newcommand{\vlsr}{\ensuremath{V_{\rm LSR}}}
\newcommand{\WJset}[1]{\mathbf{W}_{#1}}

\graphicspath{{./}{figures/}}

\shortauthors{authors}

\begin{document}

\title{A detection of H$_2$ in a high velocity cloud toward the Large Magellanic Cloud}

\author[0000-0003-0789-9939]{Kirill Tchernyshyov}
\affiliation{Department of Astronomy, University of Washington, Seattle, WA, USA}
\correspondingauthor{Kirill Tchernyshyov}
\email{ktcherny@gmail.com}

\begin{abstract}
This work presents a new detection of H$_2$ absorption arising in a high velocity cloud (HVC) associated with either the Milky Way or the Large Magellanic Cloud (LMC).
The absorber was found in an archival \emph{Far Ultraviolet Spectroscopic Explorer} spectrum of the LMC star \sknew.
This is the fifth well-characterized H$_2$ absorber to be found in the Milky Way's halo and the second such absorber outside the Magellanic Stream and Bridge.
The absorber has a local standard of rest central velocity of $+$140 km s$^{-1}$ and a H$_2$ column density of $10^{17.5}$ cm$^{-2}$.
It is most likely part of a cool and relatively dense inclusion ($T\approx 75$ K, $n_{\rm H}\sim 100$ cm$^{-3}$) in a warmer and more diffuse halo cloud.
This halo cloud may be part of a still-rising Milky Way Galactic fountain flow or an outflow from the Large Magellanic Cloud.
\end{abstract}

\section{Introduction}
\label{sec:intro}

The gaseous halos around galaxies consist mostly of diffuse, ionized gas with temperatures $T>10^{4}$ K.
They also contain some small amount of {dense} $T\sim 10 \rng 100$ K gas that can support the presence of molecular hydrogen (\molH).
It is not clear if this \molH\ typically forms in the halo itself or if it is galactic \molH\ that was ejected into the halo.
In either case, the presence of \molH\ in a cloud indicates that the cloud contains material from a galaxy: efficient \molH\ formation happens on dust grains, whose presence would be unexpected in a cloud consisting of mostly intergalactic material.
Because halo gas tends to have lower metallicities, dust-to-gas ratios, and radiation field intensities than gas in galaxies, halo \molH\ is an interesting test case for models of {the chemistry of diffuse molecular clouds.}

Halo \molH\ is detected as restframe ultraviolet absorption associated with Werner and Lyman electronic transitions of \molH.
In the Milky Way's halo, \molH\ is most often seen in {clouds with a local standard of rest line of sight velocity $\vert \vlsr \vert$\ of 20\rng90 \kms\ (intermediate velocity clouds or IVCs; \citealt{Richter:2003uf,Putman:2012vp}).}
IVCs are typically found a few kpc above the disk of the Milky Way and most likely represent gas associated with galactic fountain flows.
Analyses of gas phase elemental abundances in IVCs show that they contain dust \citep{Richter:2001vv,Werk:2019tl}.
\molH\ has also been detected in extragalactic absorbers with $\atomH$\ column densities $\NatomH \gtrsim 10^{19}$ \cmmt: damped Lyman $\alpha$ absorbers (DLAs) and sub-DLAs \citep{Levshakov:1985vs,Ledoux:2003uq,Muzahid:2015ub}.
While some of these absorbers, particularly the DLAs, may be located in galaxies, others are thought to be found in galaxy halos \citep{Muzahid:2016uc}.

Finally, there has been a small number of \molH\ detections in Milky Way {clouds with $\vert \vlsr \vert>90$ \kms\ (high velocity clouds or HVCs), which are thought to be more distant and more metal poor than IVCs.}
Three of the clear and well-characterized detections {were} found in the Magellanic system: the Leading Arm and main body of the Magellanic Stream \citep{Sembach:2001tc,Richter:2001vv} and the Magellanic Bridge \citep{Lehner:2002wz}.
The fourth detection was found in the direction of the Galactic center and may be an example of gas ejection from the Galactic disk by a nuclear wind \citep{Cashman:2021um}.

There is an additional tentative detection of \molH\ in an HVC toward the star \skold\ in the Large Magellanic Cloud (LMC; \citealt{Richter:1999vo,Bluhm:2001td,Richter:2003wo}).
However, the complexity of the stellar pseudocontinuum of \skold\ makes estimating properties of the molecular absorber infeasible.
This HVC (the \emph{HVC toward the LMC}, or HVC-L for short) is a positive velocity HVC that covers, and possibly extends beyond, the {disk} of the LMC \citep{Savage:1981te,Lehner:2009wk,Barger:2016tt}.
Despite its apparent association with the LMC, at least part of the HVC is no more than 13.3 kpc from the Sun \citep{Werner:2015vc,Richter:2015uo}.
There may be an additional structure near the LMC that appears as part of the same HVC as a projection effect \citep{Ciampa:2021up}.
The HVC-L is metal poor ($Z=0.2\rng0.4 Z_\odot$) and includes highly ionized gas \citep{Lehner:2009wk}.
A number of origins for the HVC-L have been proposed.
If some part of the HVC-L is near the LMC, that part could be a star-formation driven outflow from the LMC
\citep{Staveley-Smith:2003wo,Barger:2016tt,Ciampa:2021up}.
The part that is near the Milky Way could be infalling intergalactic medium gas or a galactic fountain flow originating in the lower-metallicity outskirts of the Milky Way \citep{Savage:1981te,Richter:2015uo}.

This work reports on a newly-discovered \molH\ absorber in the HVC-L seen toward the LMC star \sknew.
{This detection} is the fifth well-characterized HVC \molH\ absorber in the Milky Way's halo.
The data used are presented in \S\ref{sec:data}.
Analysis methods and measurements are described in \S \ref{sec:results} and discussed in \S \ref{sec:discussion}.
Finally, the results of the work are summarized in \S \ref{sec:conclusion}.

\section{Data}
\label{sec:data}

The UV observation analyzed in this work is a Far Ultraviolet Spectroscopic Explorer (\emph{FUSE}) spectrum of the LMC star \sknew\ \citep{Moos:2000vf,Moos:2002ux}.
This spectrum was recorded as part of the FUSE Legacy in the Magellanic Clouds program (PI: Blair, \emph{FUSE} PID E511, \citealt{Blair:2009wh}).
\sknew\ was observed through the MDRS aperture over a sequence of twelve exposures.
Coadded one-dimensional spectra for each of the eight \emph{FUSE} detector sides were downloaded from the Mikulski Archive for Space Telescopes.
These coadded spectra were produced {by} the archive using version 3.2.1 of CALFUSE.
A \atomH\ 21cm emission spectrum taken in the direction of \sknew\ as part the GASS survey \citep{Kalberla:2015wo} was downloaded from the Argelander-Institut f\"ur Astronomie \atomH\ Surveys Data Server\footnote{https://www.astro.uni-bonn.de/hisurvey/index.php}.
This spectrum provides a rough estimate of the column density of \atomH\ in this part of the HVC-L and serves as a velocity reference.

\begin{figure*}
    \centering
    \includegraphics[width=\linewidth]{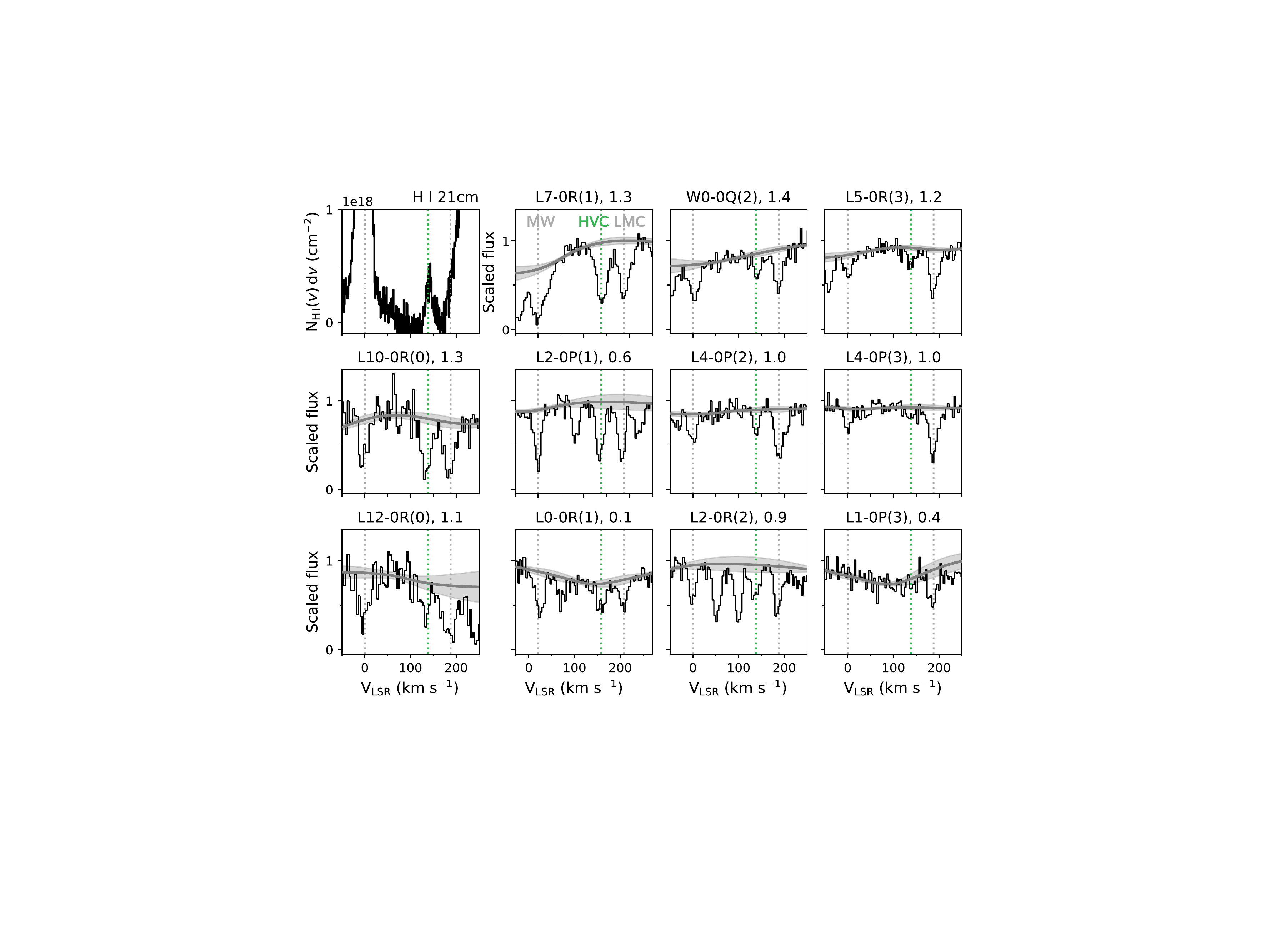}
    \caption{
    \atomH\ emission and \molH\ absorption at the velocities of the Milky Way (gray, $\approx 0$ \kms), a high velocity cloud (green, $\approx 140$ \kms), and the Large Magellanic Cloud (gray, $\approx 190$ \kms).
    The top left panel shows a \atomH\ 21 cm emission spectrum.
    The remaining panels show absorption due to different \molH\ transitions, along with some unrelated interloping absorption.
    {Velocity shifts from panel to panel of several \kms\ are the result of the limited precision of the FUSE wavelength calibration.}
    Each \molH\ panel is labeled with the transition name and the base-ten logarithm of the product $f\lambda_r$ of the oscillator strength and the rest wavelength {in Angstroms} of the featured transition.
    {The estimated stellar continuum and its 1$\sigma$ uncertainty are shown in the \molH\ panels in gray.}
    High velocity cloud \molH\ absorption is seen in all $J=0$, 1, and 2 transitions and in the stronger (i.e., higher $f\lambda_r$) $J=3$ transitions.
    {The HVC components of most of the \molH\ transitions shown here are unblended. The HVC and LMC components of the L7-0R(1) transition (top row, second from left) are  examples of mildly blended absorption features.}}
    \label{fig:sk-70d32-spectrum}
\end{figure*}

The \atomH\ 21 cm spectrum and regions of the \emph{FUSE} spectrum at the wavelengths of 11 \molH\ lines are shown in Figure \ref{fig:sk-70d32-spectrum}.
Transitions arising from the four lowest rotational levels of \molH\ are shown.
Emission and absorption are seen at three velocities: $0$ \kms, arising in the Milky Way; $140$ \kms, arising in the HVC-L; and $180$ \kms, arising in the LMC.
Absorption from the HVC component is seen in all four of the rotational levels shown.
No HVC absorption was detected from the J$\geq 4$ rotational levels.

\section{Measurements and Results} \label{sec:results}

\subsection{\molH\ column densities and Doppler parameters}
\label{sec:results:abs-params}

\molH\ column densities and Doppler parameters for the HVC component were determined by fitting a curve of growth (COG) to measurements of \molH\ line equivalent widths.
The molecular data needed for the analysis, including oscillator strengths $f$, rest wavelengths $\rstwv$, and damping constants $\gamma$, {were} taken from \citet{Abgrall:1993vh} and \citet{Abgrall:1993ux} as tabulated in the \texttt{linetools} package.

Following common practice for analyses of \molH\ in FUSE spectra, equivalent widths were measured separately for each detector segment-side combination, without coadding overlapping spectral regions \citep{Tumlinson:2002tz,Wakker:2006tj}.
In preparation for the equivalent width {measurements}, the spectra were locally continuum normalized by masking wavelength regions around the locations of absorption features and using Gaussian process regression\footnote{Done using the package \texttt{george} \citep{Ambikasaran:2015wo}.} to impute the masked continuum.
Regions were masked around the expected location of weak or undetected lines as well as detected absorption features to avoid biasing estimates of upper limits.
Gaussian process regression was done assuming a Matern $3/2$ kernel with parameters optimized to fit the unmasked continuum regions.

The equivalent widths of unblended or mildly blended \molH\ lines were measured using a combination of direct integration and Gaussian profile fitting.
A ``mildly blended" line {is one whose wings overlap the wings of another line, but whose core is unblended.}
{For example, the stronger $J=0$ and $J=1$ features of the HVC and LMC components are mildly blended at the resolution of FUSE.
This blending can be seen in the L7-0R(1) transition panel of \autoref{fig:sk-70d32-spectrum}.}
Direct integration was used for unblended lines whose absorption spanned less than 40 \kms\ or was undetected.
Stronger or mildly blended lines were measured by fitting Gaussian profiles {simultaneously to all the absorption features in the blend (if applicable).}
Gaussian profile fitting was not used for weaker lines because non-linear fits to noisy and low-contrast features are known to give measurements that are biased high \citep{Portillo:2020tm}.
{All equivalent width measurements were recorded as values with Gaussian uncertainties.
Non-detections were not converted to upper limits at this stage of the analysis.
}

The presence of \molH\ absorption in three distinct velocity components limits the number of available unblended and mildly blended transitions.
The HVC component still had a large number of usable lines arising in the J=1, 2, 3, and 4 rotational levels, but only three lines were available for the J=0 level: the Lyman 6-0, 10-0, and 12-0 R-branch transitions at 963, 981, and 1024\AA.
The HVC Lyman 10-0 transition is, technically, blended with a Milky Way Lyman 11-0 P-branch J=4 transition.
However, all other Milky Way J=4 lines were non-detections, including lines with $f\rstwv$ 5 times greater than that of the potential blend.
The contribution of the Milky Way J=4 line to the HVC J=0 absorption should therefore be negligible.

\begin{figure*}
    \centering
    \includegraphics[width=\linewidth]{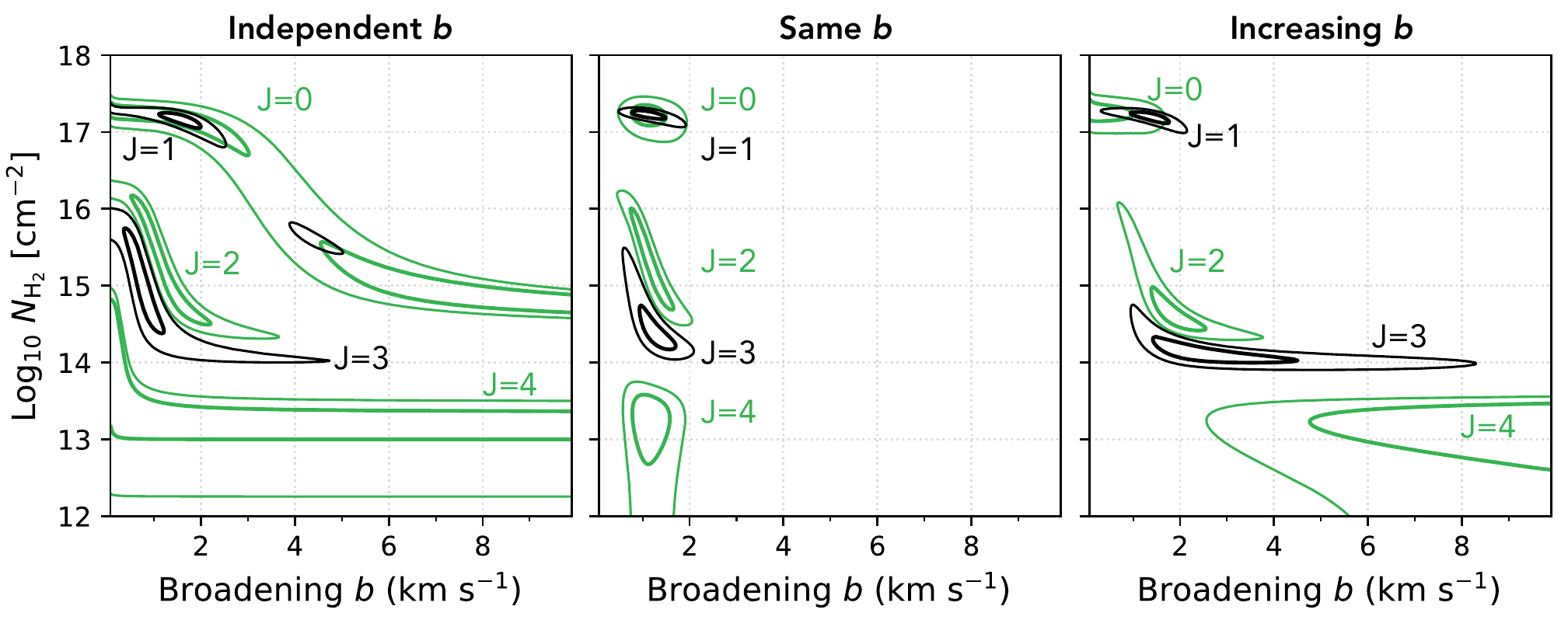}
    \caption{Curve of growth analysis of the high velocity cloud \molH\ absorption.
    Each panel shows credible regions in the Doppler parameter-column density plane.
    Inner and outer contours enclose $1 \sigma$ and $2 \sigma$ equivalent regions of the posterior probability distribution.
    The panels correspond to different assumptions for how the Doppler parameter $b$ is connected between rotational levels.
    In the left panel, each level has its own Doppler parameter with no connection to that of other levels.
    In the middle panel, the levels share a single Doppler parameter.
    In the right panel, the Doppler parameter is assumed to increase with increasing rotational level.
    }
    \label{fig:hvc-eqw-fit}
\end{figure*}

Equivalent width measurements of the same feature from different spectral segments were averaged, weighting by the inverse variance of each measurement.
A COG was then fit to the HVC equivalent width measurements {in the J=0, 1, 2, 3, and 4 rotational levels.
{The likelihood $p(W_i \vert \lten \NJ{J}, b_J)$ for an equivalent width $W_i$ measured for a transition arising in rotational level $J$ was taken to be Gaussian with mean the equivalent width of a Voigt profile of the transition with the given $\lten \NJ{J}$ and $b_J$.
The likelihood of the set of all equivalent width measurements $\WJset{J}$ arising from level $J$, $p(\WJset{J} \vert \lten \NJ{J}, b_J)$, is the product of the individual likelihoods for each $W_i$ in $\WJset{J}$.}

For each level, these likelihoods were tabulated over a grid in log column density and $b$.
The column density grid covers $\lten \, (\NJ{J}/\mcmmt)=12\rng 18$ in steps of $\Delta \lten \, (\NJ{J}/\mcmmt)=0.015$.
The $b$ parameter grid covers $b\approx0.14\rng 9.9$ \kms\ in steps of 0.0375 \kms.
The prior over each rotational level's column density, $p(\ltNJ{J})$, was taken to be uniform in logarithmic space over the range spanned by the evaluation grid.
Integrating $p(\WJset{J}\vert \ltNJ{J}, b_J) \times p(\ltNJ{J})$ over column density yields $p(\WJset{J} \vert  b_J)$, the likelihood of a level's equivalent width measurements given a value of the Doppler parameter.

Three different ways of combining information across rotational levels were used in this work.
In all three cases, the column densities of different levels were assumed to have no direct dependence on each other.
The Doppler parameters were assumed to be: (1) independent, (2) the same across levels (i.e., a single Doppler parameter $b$), or (3) increasing with increasing $J$.
Taking the Doppler parameters to be independent requires the fewest assumptions.
However, the limited line strength $f \lambda_r$ range in each rotational level leads to poorly constrained column densities.
Assuming a single Doppler parameter across levels gives a wide $f\lambda_r$ range and has been done in the \molH\ literature \citep[e.g.,][]{Tumlinson:2002tz}.
However, other \molH\ absorption analyses have shown that in some cases, the Doppler parameter increases with increasing $J$ \citep{Lacour:2005vy,Noterdaeme:2007tk,Balashev:2009tw}.
Assuming the Doppler parameter increases allows for some information sharing across levels without imposing the possibly unphysical constraint of a single Doppler parameter for all rotational levels.

These options correspond to three different priors for the level Doppler parameters.
In the independent and single Doppler parameter cases, the prior over each $b_J$ and over the single $b$ was taken to be uniform over the range spanned by the evaluation grid.
In the increasing Doppler parameter case, the Doppler parameters were taken to be a scaled and shifted cumulative sum of a vector drawn from a Dirichlet distribution.
This procedure results in a prior over vectors of increasing Doppler parameters between the minimum and maximum values of the evaluation grid.

The three cases require different computational procedures to derive a posterior probability distribution over the level column densities.
In the independent case, the posterior probability distributions (PPDs) $p(\ltNJ{J}, b_J \vert \WJset{J})$ are proportional to the level likelihoods and the univariate PPDs $p(\ltNJ{J} \vert \WJset{J})$ can be obtained by integrating the tabulated bivariate PPDs over $b_J$.
In the single Doppler parameter case, the different $\ltNJ{J}$ share a $b$ and are no longer independent.
However, they are conditionally independent given $b$.
{The PPD of $\ltNJ{J}$ and $b$ can be split into contributions from the priors, from $\WJset{J}$, and from $\WJset{J'}$\ with $J'\neq J$:}
\begin{equation}
\begin{split}
     p(\ltNJ{J}, & b \vert \{\WJset{J'}\}_{J'\neq J}, \WJset{J}) \propto p(\ltNJ{J})\, p(b)\\
     &\times p(\WJset{J}\vert \ltNJ{J}, b)
     \, \prod_{J'\neq J} p(\WJset{J'}\vert b).
\end{split}
\end{equation}
{This quantity can be calculated by combining the two-dimensional likelihood evaluation grids with the one-dimensional $p(\WJset{J'} \vert b)$ grids.
It is not necessary to first generate the joint PPD over $b$ and all five $\ltNJ{J}$. }

In the increasing Doppler parameter case, the conditional dependence structure of the model is analogous to a hidden Markov model---$b_J$ depends directly only on $b_{J-1}$, $b_{J+1}$, and $\WJset{J}$.
The bivariate PPD $p(\ltNJ{J}, b_J \vert \{\WJset{J'} \}_{J'=0, 1, 2, 3, 4})$ can therefore be calculated using a continuous-state version of the forward-backward algorithm (e.g, \citealt{Rabiner:1986wq}).
The implementation of the forward-backward algorithm for this particular problem is written out in detail in \autoref{sec:appendix}.
Briefly, the PPD over $\ltNJ{J}$ and $b_J$ can be written as the product of three terms: the likelihood of level $J$, the probability of $b_J$ given the $\WJset{J'}$ with $J'<J$, and the likelihood of the $\WJset{J'}$ with $J'>J$ given $b_J$.
The $J'<J$ term can be calculated recursively starting at $J=0$ and the $J'>J$ term can be calculated recursively starting at $J=4$.
The PPD over $\ltNJ{J}$ and $b_J$ can then be written as the product of a bivariate likelihood with the two univariate $J'\neq J$ terms.
Once again, the calculation can be done using the likelihood grids with no need to generate the joint PPD over all column densities and Doppler parameters.

{Credible regions for the three COG fits} are shown in \autoref{fig:hvc-eqw-fit} and
fit} parameters and uncertainties are listed in \autoref{tab:hvc-measurements}.
$\NJ{4}$\ is considered to be a non-detection because its $2\sigma$-equivalent uncertainty contour is consistent with the lowest $\NJ{4}$ value in the grid.
The $\NJ{J}$ detections are reported as {medians with $16 \rng 84\%$ credible intervals}.
The non-detection $\NJ{4}$ is reported as a $95\%$ upper limit.

\begin{deluxetable}{lccc}
\tabletypesize{\footnotesize}
\tablecolumns{4}
\tablewidth{0pt}
\tablecaption{Measured HVC \molH\ properties \label{tab:hvc-measurements}}
\tablehead{\colhead{Value} & \colhead{Independent $b$} & \colhead{Same $b$} & \colhead{Increasing $b$}}
\startdata
    $\lten \NmolH$ [\cmmt] & $17.2^{+0.2}_{-0.2^\dagger}$ & $17.5^{+0.1}_{-0.1}$ & $17.5^{+0.1}_{-0.1}$ \\
    $\lten \NJ{0}$ [\cmmt] & $15.5^{+1.5}_{-0.6}$ & $17.2^{+0.1}_{-0.2}$ & $17.2^{+0.1}_{-0.1}$  \\
    $\lten \NJ{1}$ [\cmmt] & $17.1^{+0.1}_{-0.4}$ & $17.2^{+0.1}_{-0.1}$ & $17.2^{+0.1}_{-0.1}$ \\
    $\lten \NJ{2}$ [\cmmt] & $15.0^{+0.9}_{-0.5}$ & $15.2^{+0.6}_{-0.5}$ & $14.6^{+0.5}_{-0.2}$  \\
    $\lten \NJ{3}$ [\cmmt] & $14.8^{+0.7}_{-0.6}$ & $14.4^{+0.5}_{-0.2}$ & $14.1^{+0.2}_{-0.1}$ \\
    $\lten \NJ{4}$ [\cmmt] & $<13.7$ & $<13.6$ & $<13.5$  \\
    $T_{01}$ (K) & $\ldots$ & $79^{+17}_{-11}$ & $74^{+14}_{-10}$ \\
    $T_{02}$ (K) & $\ldots$ & $84^{+26}_{-14}$ & $68^{+14}_{-5}$ \\
    $T_{03}$ (K) & $\ldots$ & $109^{+15}_{-7}$ & $99^{+5}_{-4}$ \vspace{0.04in}\\
\enddata
\tablecomments{Measured properties of the high velocity cloud \molH\ absorption.
Uncertainties are credible regions covering the central 68\% of each parameter's 1D posterior probability distribution.
Upper limits are {95th percentiles}. {Excitation temperatures for the independent $b$ case are not given because they are essentially unconstrained.}
{$^{\dagger}$The posterior probability distribution for the independent $b$ total column density is multimodal, with a secondary mode at lower values. The 2.5th percentile of the $\lten$\ total column density for the independent $b$ case is 15.8.}}
\end{deluxetable}

\subsection{\molH\ rotational excitation}
\label{sec:results:excitation}

\begin{figure}
    \centering
    \includegraphics[width=\linewidth]{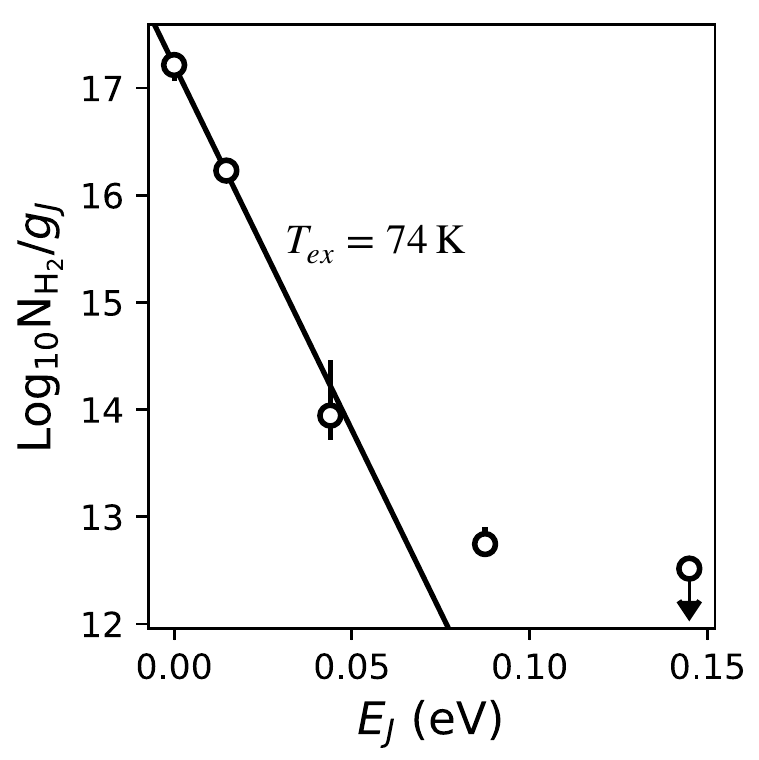}
    \caption{Excitation diagram for the rotational level populations of the high velocity cloud absorber.
    Limits and datapoints with errorbars show column densities divided by level degeneracies as a function of level energies.
    The column densities were derived under the assumption that the Doppler parameter $b$ increases with increasing $J$.
    The slope corresponding to the nominal excitation temperature between the $J=0$ and 1 levels, $T_{01}$, is shown as a black line.
    The first three rotational levels are consistent with a single temperature of about 74 K.
    The $J=3$ level requires a higher excitation temperature, indicating the influence of non-collisional excitation processes.}
    \label{fig:excitation}
\end{figure}

The population distribution of \molH\ among the rotational levels was analyzed {by calculating a series of excitation temperatures between levels using ratios of the level column densities} (e.g., $T_{01}$ between the $J=0$ and $J=1$ levels).
{The temperatures are essentially unconstrained in the independent $b$ case and are consistent within $1\sigma$ uncertainties between the same $b$ and increasing $b$ cases.}
Point estimates and uncertainties for the excitation temperatures are listed in \autoref{tab:hvc-measurements}.
{An excitation diagram with column densities from the increasing $b$ analysis is shown in \autoref{fig:excitation}.}

{The level population distribution of the \sknew\ absorber is consistent with a cool and dense cloud.
$T_{01}$ is approximately 75 K, lower than the average of $124\pm8$ K for IVCs and other high-latitude Milky Way clouds \citep{Gillmon:2006tl}.
Because the absorber's \molH\ column density is greater than $10^{16}$ \cmmt, $T_{01}$ is likely to be close the gas kinetic temperature \citep{Roy:2006vn}.
In both the independent and increasing $b$ cases, $T_{01}$ and $T_{02}$ are consistent with each other while $T_{03}$ is greater than $T_{01}$, meaning that the \molH\ level populations up to and including the $J=2$ level are thermalized while the $J=3$ level and above are not.
The volume density of the gas is therefore likely to be between the critical densities for these two levels, $\nmolH \sim 200-3000$ \dens \citep{Jorgenson:2010vu}.
}

\subsection{\atomH\ column density}
\label{sec:results:atomH}
A direct measurement of the HVC component's \atomH\ column density is not possible because the HVC component's UV $\atomH$\ absorption is blended with absorption from the stronger Milky Way and LMC components.
{Instead}, \NatomH\ has to be estimated through indirect methods.
{One} method is to use the 21 cm emission spectrum in the direction of the absorber.
This provides an \NatomH\ measurement for a region that includes the sightline, but also includes emission from surrounding gas.
{A} second method is to combine the HVC's estimated metallicity with a measurement of the \specnotation{O}{I} column density.
The 21 cm emission method gives a total column density of $10^{18.85}-10^{19.15}$ \cmmt.
The \specnotation{O}{I} column density toward \sknew\ is $10^{15.35}$ \cmmt\ and the metallicity is $0.2\rng0.4$ times the solar metallicity (as defined by \citealt{Lodders:2009ww}), giving $\NatomH=10^{19}-10^{19.3}$ \cmmt \citep{Lehner:2009wk}.
Combining the ranges produced by the two methods yields $\NatomH = 10^{18.85}-10^{19.3}$ \cmmt.

\subsection{Molecular fraction}
\label{sec:results:fmolH}

\begin{figure*}
    \centering
    \includegraphics[width=\linewidth]{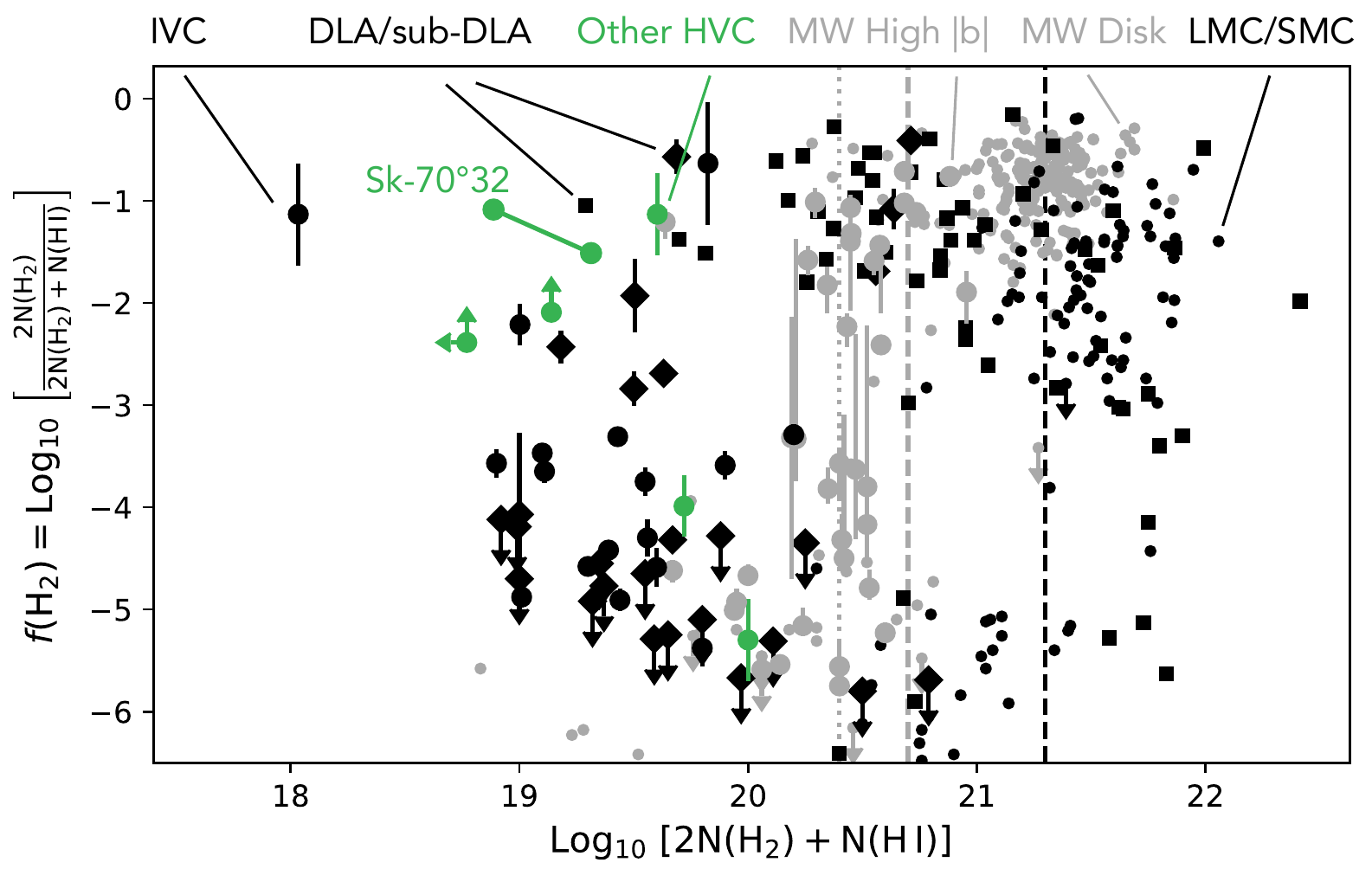}
    \caption{Molecular fractions \fmolH\ as a function of total hydrogen column \NH\ for different populations of molecular absorbers.
    The populations are found in the Magellanic Clouds (small black circles; \citealt{Welty:2012vl}), the Milky Way disk at low and high Galactic latitudes (small and large gray circles; \citealt{Savage:1977uc,Gillmon:2006tl,Rachford:2009vw,Shull:2021wr}), intermediate velocity clouds (large black circles; \citealt{Richter:2003wo,Wakker:2006tj}), high velocity clouds including the \sknew\ absorber (green circles; \citealt{Richter:2001vv,Sembach:2001tc,Lehner:2002wz,Wakker:2006tj,Cashman:2021um}), and extragalactic Damped and sub-Damped Lyman $\alpha$ absorbers at $z<0.6$ (black diamonds; \citealt{Muzahid:2015ub,Muzahid:2016uc}) {and $1.8<z$ (black squares; \citealt{Noterdaeme:2008te,Balashev:2014vc,Ledoux:2015vo,Noterdaeme:2018wi,Balashev:2019wa,Ranjan:2020we}. }
    The two connected \sknew\ datapoints represent the uncertainty on the total column density of the absorber.
    Transition \NH\ values between low and high \fmolH\ are shown as vertical lines for the Milky Way high latitude (dotted gray), Milky Way disk (dashed gray), and Magellanic Cloud (dashed black) populations.}
    \label{fig:fH2}
\end{figure*}

The molecular fraction is the fraction of H atoms that are in the form of \molH.
Assuming that the amount of ionized hydrogen in the molecular gas is negligible, the {local} molecular fraction {at a point along a sightline is $\frac{2 \nmolH}{2\nmolH+\natomH}$.
This local quantity will vary with depth into the molecular gas}.
{Taking the $\nH$-weighted average of the local molecular fraction gives the (sightline-averaged) molecular fraction $\fmolH \equiv \frac{2 \NmolH}{2\NmolH+\NatomH}$.}
The denominator of this fraction is the total un-ionized hydrogen column density, $\NH$.
The molecular fraction of the HVC component is 0.03\rng0.08, where the uncertainty is dominated by the uncertainty in $\NatomH$.
\autoref{fig:fH2} shows the molecular fraction as a function of total un-ionized hydrogen column density for the \sknew\ HVC absorber and for sightlines in the Milky Way disk, the LMC and SMC, low redshift DLAs and sub-DLAs, high-latitude Milky Way sightlines including IVCs, and other HVCs.

The molecular fraction is set by the balance between \molH\ formation and dissociation.
{Because \molH\ formation happens most efficiently on dust grain surfaces, the \molH\ formation rate depends on metallicity via the dust-to-gas ratio.
The metallicity of the HVC-L has been measured to be 0.2-0.4 times the solar metallicity, meaning that the absorber should have a \molH\ formation rate that is several times lower than the rate at solar metallicity.
Dissociation can happen through collisions or through photodissociation by UV photons.
The absorber's relatively low \molH\ excitation temperature suggests that the gas has not recently experienced a fast shock \citep{Wilgenbus:2000up}.
The \molH\ fraction therefore depends mostly on the gas density and the radiation field strength, though the dependence on the radiation field strength is not linear because the \molH\ column density is high enough for self-shielding to be important.
Comparing the location of the \sknew\ absorber with other systems shown in \autoref{fig:fH2}, its \fmolH\ is higher than is typical for its \NH.
Given the lower than solar formation rate, the high \fmolH\ suggests that the absorber is particularly dense or that the radiation field strength at the absorber's location is particularly weak.
}

\subsection{Physical conditions}
\label{sec:results:conditions}
With a few assumptions, the density \nH\ and incident radiation field strength $I_{UV}$ can be estimated from the column densities of \atomH\ and the \molH\ rotational levels.
In this work, this was done by generating models of clouds with different \nH\ and $I_{UV}$ and comparing the model and observed column densities.
Qualitatively, this comparison combines two constraints: the molecular fraction \fmolH\ and the excitation of the non-thermalized higher-J rotational levels of \molH\ (e.g., \citealt{Jura:1975uy,Lee:2007tx,Klimenko:2020vt}).

Models were generated using the \texttt{Cloudy} photoionization code \citep{Ferland:2017wx}\footnote{Version 17.02} with the \citet{Shaw:2005vj} \molH\ implementation.
The molecular cloud is assumed to be a plane-parallel slab with a single density \nH\ and a constant temperature.
The cloud is illuminated by the cosmic microwave background and by a scaled \citet{Draine:1978vc} radiation field.
The cloud metallicity is set to 0.3 times solar, the nominal metallicity determined by \citet{Lehner:2009wk}.
The dust-to-gas ratio is set to the solar value scaled by the metallicity, i.e., assuming a fixed dust-to-metals ratio.

\begin{deluxetable}{lcc}
\tabletypesize{\footnotesize}
\tablecolumns{3}
\tablewidth{0pt}
\tablecaption{\texttt{Cloudy} model grid parameters \label{tab:cloudy-grid}}
\tablehead{\colhead{Parameter} & \colhead{Range} & \colhead{Stepsize}}
\startdata
$\lten (I_{UV}/\nH)$ [cm$^3$] & $-2.3$ to $-2.2$ & 0.05 \\
$\lten \nH$ [\dens] & $2$ to $3$ & 0.2 \\
$T$ (K) & 70 to 110 & 20 \\
\enddata
\tablecomments{Parameters varied to generate a grid of \texttt{Cloudy} models. $I_{UV}$ is the amplitude of the \citet{Draine:1978vc} field.}
\end{deluxetable}

Models were generated at points over a grid in $\lten \nH$, $\lten (I_{UV}/\nH)$ (where $I_{UV}$ is in units of the \citet{Draine:1978vc} field), and temperature.
After an initial exploration over a broad and coarse grid in these parameters, the more localized and refined grid listed in \autoref{tab:cloudy-grid} was used.
The logarithmic column densities at the grid points were then interpolated to a grid fine enough to resolve the $\lten\NJ{J}$\ {PPDs}.

{Comparisons between the models and observations were done separately using the same $b$ and increasing $b$ column density PPDs.
The results for the two calculations overlap, but do not identically agree.
Taking the two cases to be equally likely yields an estimated $\nH=$ 100 to 500 \dens\ and a radiation field that is 0.3 to 1.6 times the \citet{Draine:1978vc} field.}
These ranges reflect the uncertainty on the $\lten\NJ{J}$\ measurements, but do not include systematic uncertainties such as the unknown true cloud geometry and dust-to-metals ratio.

\section{Discussion} \label{sec:discussion}

\subsection{The location and nature of the \sknew\ HVC absorber}
\label{sec:discussion:location-and-nature}

\begin{figure*}
\includegraphics[width=\linewidth]{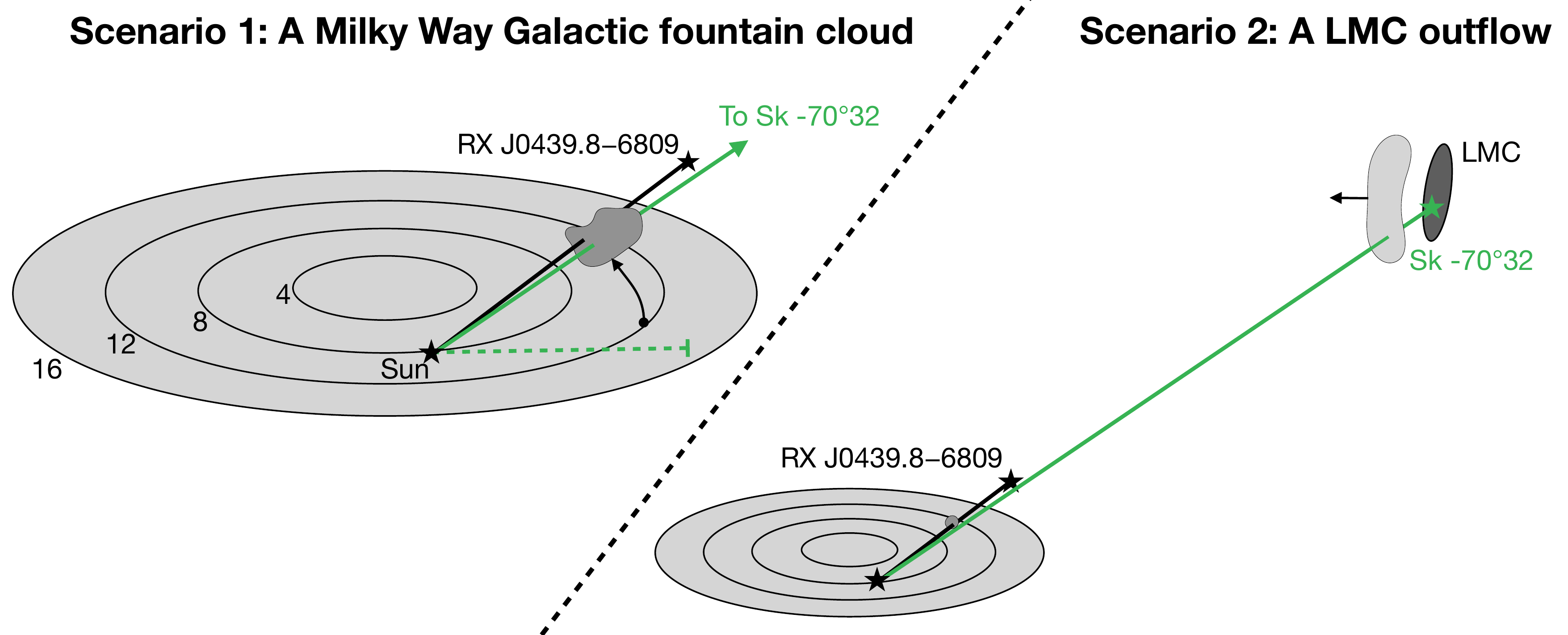}
\caption{Two possible origin scenarios for the \sknew\ molecular absorber: a Milky Way Galactic fountain flow (left, scenario 1) and a Large Magellanic Cloud outflow (right, scenario 2). In scenario 1, the \sknew\ absorber is  part of a Milky Way HVC that is seen toward the star RX J0439.8-6809 \citep{Werner:2015vc,Richter:2015uo}. In the second scenario, the \sknew\ absorber is part of an LMC outflow and the Milky Way HVC seen toward RX J0439.8-6809 is a chance alignment.}
\label{fig:diagram}
\end{figure*}

There are three possible {origin scenarios} for the HVC-L molecular absorber: an inflow originating in the intergalactic medium (IGM), a Milky Way galactic fountain cloud, and an LMC outflow; the two outflow scenarios are shown in \autoref{fig:diagram}.
The presence of \molH\ at relatively high \fmolH\ in the cloud argues against an IGM inflow.
Efficient \molH\ formation requires dust grain surfaces, while an IGM inflow would contain little to no dust.
Both outflow scenarios are possible, but both come with tensions.
A Milky Way galactic fountain cloud would be kinematically extreme, while an LMC outflow would require the HVC-L to be a coincidental on-sky alignment of two physically unrelated HVCs.

If the absorber is part of a Milky Way galactic fountain flow, it should be within 13.3 kpc of the Sun \citep{Werner:2015vc,Richter:2015uo} and its rotational velocity about the Galactic center $V_\phi$ should be between $V_\phi$ of gas at the flow's origin point and at its current height.
The measured lag in the rotational velocity of extraplanar \atomH\ as a function of height off the plane is ${-15}$ \kms\ kpc$^{-1}$ \citep{Marasco:2011uv}, so at the upper bound on the distance to the cloud $V_\phi\approx100$ \kms.
Assuming that the absorber's $V_\phi$ is greater than or equal to the lagged $V_\phi$ at its height, its measured $\vlsr$\ requires the cloud to be some combination of (1) at least 4 kpc away, (2) moving away from the plane, and (3) moving outward away from the Galactic center.

\begin{figure*}
\includegraphics[width=\linewidth]{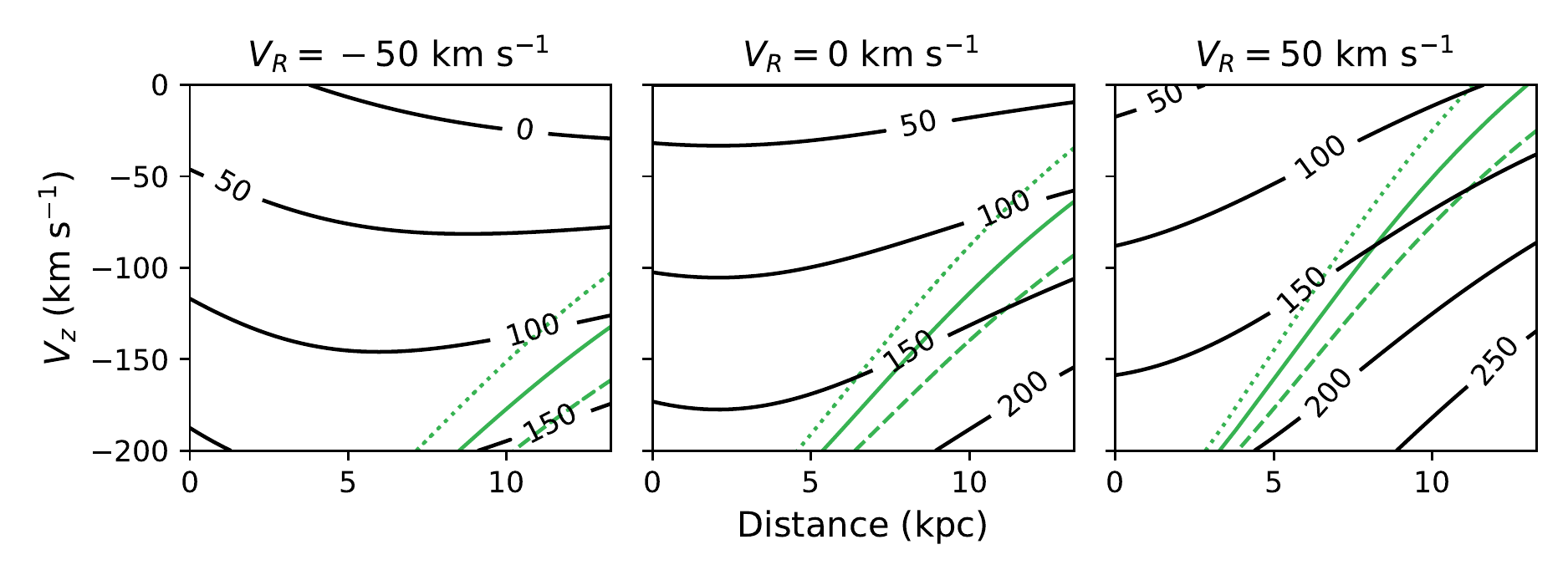}
\caption{Combinations of distance to the \sknew\ absorber and values of different Galactocentric cylindrical velocity components that match the observed \vlsr\ of the high velocity cloud absorber, assuming the absorber is a Milky Way galactic fountain cloud (Scenario 1 in \autoref{fig:diagram}).
Panels correspond to different values of the component toward or away from the Galactic center, $V_R$.
Contours correspond to different values of the rotational component $V_\phi$.
Curves corresponding to constant lags in $V_\phi$ with height off the plane of $-11$, $-15$, and $-19$ \kms\ kpc$^{-1}$ (dashed, sold, and dotted lines) are shown in green.}
\label{fig:velo-matching}
\end{figure*}

Figure \ref{fig:velo-matching} shows different possible combinations of distance, vertical velocity $V_z$, cylindrical radial velocity $V_R$, and $V_\phi$ that agree with the absorber's measured line-of-sight velocity.
At the nominal distance of the constraining measurement from \citet{Richter:2015uo}, the absorber would be 5.3 kpc below the Galactic plane.
Assuming the most favorable $V_R$ shown, $+50$ \kms, the absorber would have $V_Z\approx-70$ \kms; a $V_R$ near 0 \kms\ would require $V_Z\approx-130$ \kms.

For comparison, the galactic fountain flow proposed by \citet{Marasco:2017wq} as an explanation for the Smith Cloud has a vertical velocity of less than 75 \kms\ away from the plane at a height of 3 kpc.
\citet{Marasco:2017wq} note that the energy required to launch the cloud on this trajectory is high, though still plausible.
The energy required to produce a cloud with the kinematics of the HVC-L absorber would presumably be even more extreme.

If the absorber is instead part of an outflow from the LMC, there would need to be at least two physically distinct but observationally similar HVCs in this part of the sky: one associated with the LMC and one within 13.3 kpc associated with the Milky Way \citep{Richter:2015uo}.
As \citet{Ciampa:2021up} argue, this coincidence would not be extreme given the incidence rate of compact HVCs.
The existence of an LMC outflow at the HVC-L's velocity range is supported by {several pieces of}  circumstantial evidence, including the observation of a corresponding redshifted gas component in spectra taken toward sources behind the LMC but not sources in the LMC itself \citep{Barger:2016tt}.
In this scenario, the HVC-L {could} be a less-molecular LMC analogue to the outflow found off the Small Magellanic Cloud by \citet{Di-Teodoro:2019vl}.

In both scenarios, the HVC-L would have been ejected from its origin galaxy with a substantial initial velocity.
Again taking the \citet{Marasco:2017wq} Smith Cloud model as a reference, the initial velocity in the Milky Way galactic fountain flow scenario would have been $\approx 185$ \kms\ or greater.
In the LMC outflow scenario, the velocity offset between the \sknew\ absorber and the LMC bulk velocity in that direction is $\approx 100$ \kms.
The {velocity of the LMC relative to the Milky Way's halo} would mean that this flow is encountering a headwind of around 200 \kms\ along the direction perpendicular to the LMC's disk.
Depite these launch velocities and headwinds, the HVC-L contains pockets of cool ($T\approx75$ K) and dynamically quiescent ($b\approx 1$ \kms\ for $J=0$) gas.

\subsection{The \sknew\ HVC absorber on the sequence of \molH\ transitions}
\label{sec:discussion:context}

Well-defined populations of \molH\ absorbers show evidence of an atomic-to-molecular transition in the \NH-\fmolH\ plane: there exists a value of \NH\ that divides most sightlines with $\fmolH\lesssim 1$\% and $\fmolH\gtrsim 1$\%.
The transition point is set by the balance between {the} radiation field strength {and} the \molH\ formation rate (e.g., \citealt{McKee:2010tc}).
The transition point is at $\lten \NH/$\cmmt$ \approx21.3\rng22$ in the Magellanic Clouds \citep{Tumlinson:2002tz,Welty:2012vl}, 20.7 in the Milky Way disk at low Galactic latitudes \citep{Savage:1977uc,Shull:2021wr}, and 20.4 in Milky Way disk clouds at high latitudes \citep{Gillmon:2006tl}.

Halo absorbers---IVCs, HVCs, and extragalactic sub-DLAs---do not have an obvious transition point, but do occupy a part of the \NH-\fmolH\ plane that is devoid of in-galaxy absorbers.
\autoref{fig:fH2} shows \fmolH\ as a function of \NH\ for different in-galaxy and halo populations.
At $\lten \NH/$\cmmt$<20$ and $\lten \fmolH>-3$, there is only one in-galaxy absorber but multiple halo absorbers, including the \sknew\ absorber discussed in this work.
{This difference in the $\lten \NH$-$\fmolH$\ distribution indicates that} the radiation field at the distances of halo clouds is weak enough to offset the typically lower metallicities and dust-to-gas ratios relative to in-galaxy absorbers.
The lack of a {distinct} transition $\lten \NH$\ may reflect a greater range in metallicities and radiation field strengths among halo clouds relative to in-galaxy clouds.

The \sknew\ absorber lies on the upper envelope of the distribution of halo absorbers in the \NH-\fmolH\ plane.
Compared with other halo absorbers, it also has a lower-than-typical $J=0\rng1$ excitation temperature.
A fit to the locus of high latitude points in \citet{Wakker:2006tj} predicts a temperature of 130 K for an absorber with $\NmolH=10^{17.5}$ \cmmt, but the measured temperature is $\approx 75$ K.
The occupation ratios of the higher $J$ levels for the \sknew\ absorber are uncertain, but lie on the lower end of what is seen in \citet{Wakker:2006tj}.
{This difference would suggest a lower degree of radiative excitation, which could be explained by the \sknew\ absorber being at a greater height off the Milky Way or LMC than other halo absorbers.}
Alternatively, if the absorber is part of the Milky Way galactic fountain and is still rising off the plane, it may contain more disk material than a typical halo cloud.

\section{Conclusion} \label{sec:conclusion}
This work presents a new detection of \molH\ absorption in a Milky Way HVC toward the LMC.
{The} absorption was found in an archival \emph{FUSE} spectrum of the LMC star \sknew.
The absorber's rotational level column densities and Doppler parameters were measured from this spectrum using a curve of growth analysis; the total \NmolH\ was found to be $10^{17.5}$ \cmmt.

The absorber could be part of a Milky Way galactic fountain flow or part of a LMC outflow.
However, its central velocity would require the galactic fountain flow to have been launched with an exceptionally high initial velocity.
The absorber has a \molH\ fraction $\fmolH$\ of 0.03\rng0.08, a rotational temperature $T_{01}\approx 75$ K, and a $J=0$ Doppler parameter $b\approx 1$ \kms, suggesting a cool and quiescent environment.
A comparison of the rotational level column densities with a grid of \texttt{Cloudy} models suggests that the absorbing cloud has a density of order $10^2$ \dens\ and is illuminated by a radiation field that is similar in strength to the \citet{Draine:1978vc} field.

This detection is the fifth well-characterized Milky Way HVC molecular absorber and is currently one of two such absorbers not found in the Magellanic Stream or Bridge.
The \sknew\ absorber is 2.69 degrees away from an HVC \molH\ detection toward \skold, for which characterization has not been possible \citep{Richter:2003wo}.
This angular separation would correspond to a physical separation of 235 pc at a distance of 5 kpc or 1409 pc at a distance of 30 kpc.
The two absorbers have similar velocities and may be part of the same cloud complex.
An examination of a total of 67 \emph{FUSE} spectra in the direction of the LMC revealed no HVC \molH\ absorption toward any other background source, a covering fraction of 2-6\%.
This can be compared with the covering fraction found for IVCs, 38-54\% \citep{Wakker:2006tj}.
The non-detections include four sources that are within 30 arcminutes (44 and 262 pc at 5 and 30 kpc) of \sknew.
The overdensities associated with the two \molH\ detections are therefore likely to be distinct local density maxima rather than different locations within a single density peak.

\begin{acknowledgements}
KT thanks Jess Werk for providing comments on a version of this manuscript, Chris Howk for useful discussions, and the anonymous referee for providing an informative and helpful report.
This work was done with support from NSF-AST 1812521, the Research Corporation for Science Advancement, Cottrell Scholar grant ID number 26842, and program \#HST-AR-16635 provided by NASA through a grant from the Space Telescope Science Institute.
\end{acknowledgements}

\software{\texttt{astropy} \citep{Astropy-Collaboration:2013uv,Astropy-Collaboration:2018vm},
\\ \texttt{emcee} \citep{Foreman-Mackey:2013ux},
\\ \texttt{george} \citep{Ambikasaran:2015wo},
\\ \texttt{linetools} \citep{Prochaska:2017vh}, 
\\ \texttt{matplotlib}  \citep{Hunter:2007ux}, 
\\ \texttt{numpy} \citep{Harris:2020ti}, 
\\
\texttt{pandas} \citep{McKinney:2010vw}
}

\appendix
\section{Details of the increasing Doppler parameter model}
\label{sec:appendix}
In the increasing $b$ parameter COG model introduced in \S\ref{sec:results:abs-params}, the column densities and Doppler parameters of the different levels depend on each other.
The resulting inference problem involves $2(J_{\rm max} + 1)$ parameters, where $J_{\rm max}$ is the highest rotational level included in the COG analysis.
If the prior over the set of Doppler parameters can be factorized as a sequence of conditional distributions,
\begin{equation}
\label{eqn:one-ahead}
p(b_0, b_1, \ldots, b_{J_{\rm max}}) =
p(b_0)\, \prod_{J=0}^{J_{\rm max} - 1} p(b_{J+1}\vert b_J),
\end{equation}
the posterior probability distributions for each level's $\ltNJ{J}$ and $b_J$ can be calculated without first generating the full $2(J_{\rm max}+1)$-dimensional posterior probability distribution of the complete model.

The target quantity is $p(b_{J}, \ltNJ{J} \vert \{\WJset{J'}\})$.
$\WJset{J}$ is the set of equivalent width measurements for level $J$ and $\{\WJset{J'}\}$ represents the collection of all the analyzed levels' equivalent width measurement sets.
The likelihood $p(\WJset{J} \vert b_J, \ltNJ{J})$ and the likelihood marginalized over the column density, $p(\WJset{J} \vert b_J)$, are described in \S\ref{sec:results:abs-params}.
The prior over the set of Doppler parameters is derived from the Dirichlet distribution:
\begin{equation}
\begin{split}
x_0, x_1, \ldots, x_{J_{\rm max}}, x_{J_{\rm max}+1} &\sim \text{Dir}(\vec{\alpha}=\mathbf{1})\\
\beta_J &= \sum_{i=0}^{J} x_J\\
b_J &= b_{\rm min} + \beta_J \times (b_{\rm max} - b_{\rm min}).
\end{split}
\end{equation}
The Dirichlet distribution over $K$ dimensions is defined over the $K-1$ dimensional simplex.
Each $x_J$ takes on a value between 0 and 1.
The $\beta_J$ are a cumulative sum of the $x_J$ and so are increasing and take on values between 0 and 1, with $\beta_{J_{{\rm max}+1}} \equiv 1$.
This last fact is the reason for using the $J_{{\rm max}+2}$ dimensional Dirichlet distribution to produce a prior over $J_{{\rm max}+1}$ variables.
The vector of concentration parameters $\vec{\alpha}$ determines the shape of the distribution, with a vector of all ones corresponding to a uniform distribution over the simplex.

As is required by \autoref{eqn:one-ahead}, the prior on the Doppler parameters can be written as a sequence of conditional distributions.
This factorization is done using a "string cutting" or "stick breaking" representation of the variable generation process:
\begin{equation}
\begin{split}
\phi_J &\sim \text{Be}(1, J_{\rm max}+1-J)\\
\beta_0 = x_0 &= \phi_0\\
x_{J+1} &= (1-\beta_{J}) \phi_{J+1}\\
\beta_{J+1} &= \beta_{J} + x_{J+1}.
\end{split}
\end{equation}
The $\phi_J$ variables are drawn from a beta distribution, represent the fraction of the still-unassigned part of the string/stick that gets assigned to $x_J$, and are independent of each other.
The conditional probability of $b_{J+1}$ given $b_J$ is proportional to that of $\beta_{J+1}$ given $\beta_J$, which can be written in terms of $\phi_{J+1}$:
\begin{equation}
\label{eqn:incr-prior}
\begin{split}
p(b_{J+1} \vert b_J) \propto p(\beta_{J+1} \vert \beta_{J}) &= p(\phi_{J+1})
\frac{{\rm d} \phi_{J+1}}{{\rm d} \beta_{J+1}}\\
&= \frac{\text{Be}(\phi_{J+1}; 1, J_{\rm max} - J)}
{1-\beta_J}
\end{split}
\end{equation}

The posterior probability distribution over $\ltNJ{J}$ and $b_J$ can be split into three terms:
\begin{equation}
    p(b_{J}, \ltNJ{J} \vert \{\WJset{J'}\}) \propto
    p(\WJset{J}\vert b_J, \ltNJ{J}) \,
    p(\ltNJ{J}) \,
    p(b_{J} \vert \{\WJset{J'}\}_{J'\neq J}).
\end{equation}
These terms are the likelihood for level $J$, the prior over $\ltNJ{J}$, and the dependence of $b_J$ on the other levels.
The last of these can be evaluated using the forward-backward algorithm, which further splits the expression into a part that depends on levels with lower $J$ (the forward contribution) and a part that depends on levels with higher $J$ (the backward contribution):
\begin{equation}
   p(b_{J} \vert \{\WJset{J'}\}_{J'\neq J}) \propto
   p(b_{J} \vert \{\WJset{J'}\}_{J'<J}) \,
   p(\{\WJset{J'}\}_{J'>J} \vert b_J).
\end{equation}
These two parts can be evaluated recursively.

The forward contribution is evaluated starting at $J=0$, where $p(b_{0} \vert \{\WJset{J'}\}_{J'<0})$ is simply the prior, $p(b_0)$.
The forward contribution for level $J>0$ is an integral involving the prior from \autoref{eqn:incr-prior} and the previous level's likelihood and forward contribution:
\begin{equation}
\label{eqn:ascending}
    p(b_{J} \vert \{\WJset{J'}\}_{J'<J}) \propto
    \int_{b_{\rm min}}^{b_{\rm max}} p(b_{J} \vert b_{J-1}) \,
    p(\WJset{J-1} \vert b_{J-1})\,
    p(b_{J-1} \vert  \{\WJset{J'}\}_{J'<J-1})\,
    {\rm d} b_{J-1}.
\end{equation}
This integral can be done numerically using tabulated likelihoods and forward contributions for the previous level.

The backward contribution is evaluated starting at $J_{\rm max}$, where it is undefined and can be taken to be unity.
For $J<J_{\rm max}$, the backward contribution is an integral similar to that of the forward contribution:
\begin{equation}
\label{eqn:descending}
    p(\{\WJset{J'}\}_{J'>J} \vert b_J) \propto
    \int_{b_{\rm min}}^{b_{\rm max}}
    p(b_{J+1} \vert b_{J})\,
    p(\WJset{J+1}\vert b_{J+1})\,
    p(\{\WJset{J'}\}_{J'>J+1} \vert b_{J+1})\,
    {\rm d} b_{J+1}.
\end{equation}
As with the forward contribution, the integral can be done numerically using tabulated quantities.
Finally, the forward, backward, and $J$-level contributions are combined to obtain $p(b_{J}, \ltNJ{J} \vert \{\WJset{J'}\})$.

\bibliography{main}
\end{document}